\documentclass[aps,reprint,twocolumn,groupedaddress,superscriptaddress]{revtex4-1}
\usepackage{graphicx}
\usepackage{amsmath,amssymb}
\usepackage{color}
\usepackage{textcomp}
\usepackage{float}

\begin{document}

\begin{titlepage}
\title{Numerical method for computing the free energy of glasses}
\author{H. A. Vinutha}
\affiliation{Institute of Physics, Chinese Academy of Sciences, Beijing, China}
\affiliation{Department of Chemistry, University of Cambridge, Cambridge, UK}
\author{Daan Frenkel}
\affiliation{Department of Chemistry, University of Cambridge, Cambridge, UK}

% word limit 600; currently 292
\begin{abstract}
{\textcolor{black}{We propose a numerical technique to compute the equilibrium free energy of glasses that cannot be prepared quasi-reversibly.  
For such systems,  standard techniques for estimating the free energy by extrapolation, cannot be used.
Instead, we use a procedure that samples the equilibrium partition function of the basins of attraction of the different inherent structures (local potential energy minima) of the system.
If all relevant inherent structures could be adequately sampled in the (supercooled) liquid phase, our approach would be rigorous. 
In any finite simulation, we will miss the lower-energy inherent structures that become dominant at very low temperatures. We find that our free energy estimates for a Kob-Andersen glass are lower than those obtained by very slow cooling,  even at temperatures down to one third of the glass transition temperature.}  

 \textcolor{black}{The current approach could be applied to compute the chemical potential of ultra-stable glassy materials, and should enable the  estimation of their solubility.}}
\end{abstract}
\maketitle
\end{titlepage}
%Intro 
\section{Introduction}
\textcolor{black}{Most glasses are prepared in a state that is far away from equilibrium, and many of the interesting properties of glasses, such as ageing, are related to this fact \cite{cavagna2009supercooled,debenedetti1996metastable,o2003jamming}. 
However, there are also examples of glassy materials that can be prepared, either experimentally \cite{singh2013ultrastable} or computationally \cite{berthier2019efficient}, in an ultra-stable state.  
The properties of such ultra-stable glasses are of great interest, not just because they offer insight in the nature of the glassy state, but also because experimentally prepared ultra-stable glasses could have interesting properties,  such as low solubility.
However, the experimental procedures by which ultra-stable glasses are prepared are less suited to determine their free energy, and thus to quantify their thermodynamic stability. Moreover, the computational techniques to prepare ultra-stable glasses are limited to systems that can be equilibrated using ``swap moves''\cite{berthier2019efficient,tsai1978structure}. For most glassy materials, such an approach will not work. Here we investigate to what extent the free energy of stable glasses can be computed starting from instantaneously quenched, extreme non-equilibrium states.}

Many aspects of the dynamics of supercooled liquids can be understood in terms of the underlying  potential energy landscape (PEL)~\cite{debenedetti2001supercooled,stillinger1995topographic}. 
A configuration of the system is represented by a point on this $ND$ dimensional hypersurface, where $N$ denotes the number of particles in the system and $D$ its spatial  dimensionality. 
The PEL has a complicated topography comprising determined numerous potential energy minima or ``inherent structures'' (IS) separated by barriers. 
All configurations of the system that, upon energy minimization,  uniquely map onto a given IS form the {\em basin of attraction} of that IS. 
At high temperatures ($T$), the thermal energy of the system is sufficient to cross the barriers separating the different basins and, as a consequence, the system can explore all allowed configurations.
However, at low $T$,  the system typically samples the configurations in one basin for a long time, and infrequently hops to another basin. 
\textcolor{black}{An energy minimization from a (supercooled) liquid state, will not populate the different inherent structures with a probability corresponding to their low-temperature Boltzmann weight~\cite{stillinger1995topographic,sciortino2005potential}.
To prepare a low temperature ``equilibrium'' glass, three distinct conditions should be met:
 first of all, the system should be able to relax the degrees of freedom associated with its motion {\em within} the potential-energy basin of a given IS.
 Secondly, the system should be able to reach the low-temperature equilibrium distribution  {\em between} the basins of attraction the different inherent structures that have been reached by quenching.
 Finally, all relevant low-temperature basins should be reachable by quenching from higher temperature. } 

\textcolor{black}{Computationally, equilibration within a basin is typically not a problem, but the other two equilibration  processes are not properly accounted for in conventional simulations. In what follows, we sketch an approach that makes it possible to achieve complete equilibration between all IS that can be sampled at high temperatures. 
Using this approach, we can estimate the free energy of the low temperature glass - but this estimate would only be rigorous if all relevant IS can be sampled in the (supercooled) liquid phase. 
However, it is well known that this is not the case for glasses at very low temperatures~\cite{sciortino2005potential,heuer2008exploring}. 
Hence, the question that we have to address is: how serious is the inadequate sampling of low-energy inherent structures for the estimate of the free energy of a stable, but not ultra-cold glass ?}

\textcolor{black}{It is important to emphasize the difference between our approach and earlier work. 
Several authors have computed (or estimated) the free-energy of individual basins in the context of estimating the configurational entropy of a glass~\cite{sastry2000evaluation,la2003numerical,sastry2001relationship,ozawa2018configurational}.
However, such an approach can only work if the free energy of the low temperature glass is already known.  
The key problem with these approaches is that the standard methods to compute (or estimate) the free energy of the basin of an IS at low temperatures, fail at high temperatures.
In the present paper, we solve this problem. }

%%%%%%%%%%%%
\subsubsection{Method}
Our method is based on the observation that the basin of attraction of an inherent structure is well defined at any temperature: in particular, there is no ambiguity in the definition of the partition function of such a basin.
Moreover, upon cooling a system constrained to be in such a basin, the system cannot fall out of equilibrium, because a basin has only a single potential energy minimum, hence trapping in local minima is excluded. 
The second key point where we differ from existing approaches is that we do {\em not} compute the free energy of the various basins, but only the free-energy difference between a given basin at high and low temperatures.  

As the configuration space of the system can be decomposed uniquely into discrete basins (labelled by $i$), we can write the low-temperature partition function of the system as 
\begin{equation}
Q(T_L)=\sum_i q_B^i(T_L) \;.
\end{equation}
We can trivially rewrite this expression as
\begin{equation}\label{eq:QL_from_QH}
Q(T_L)=\sum_i q_B^i(T_L) = Q(T_H)\times \sum_i \left({q_B^i(T_H)\over Q(T_H)}\right)\left( {q_B^i(T_L)\over q_B^i(T_H)}\right)
\end{equation}
The crucial point is that the ratio   $q_B^i(T_H)/Q(T_H)$  is simply the probability $P_i$ that basin $i$ is sampled (at $T_H$).
Hence, in a Monte Carlo sampling at $T_H$, \textcolor{black}{we visit the $i^{th}$ basin $m_i$ times with the probability $P_i=\left\langle m_i/M\right\rangle$,} where $M$ denotes the number of  distinct MC basin samples (typically a few hundred to a few thousand). Then:
\begin{equation}\label{eq:QL_from_QH2}
Q(T_L)= Q(T_H)\times \left\langle {q_B^i(T_L)\over q_B^i(T_H)}\right\rangle_{MC}
\end{equation}
or
\begin{equation}\label{eq:FL-FH}
\beta_L F_L = \beta_H F_H  -\ln \left\langle {q_B^i(T_L)\over q_B^i(T_H)}\right\rangle_{MC}
\end{equation}
For a given basin, ${q_B^i(T_L)\over q_B^i(T_H)}$ can be computed by normal thermodynamic integration (by construction, a system inside a single basin cannot be trapped in local minima).
Note that, in computing the basin partition function, we do not make use of any  harmonic or quasi-harmonic approximation. 
The method that we use to reject trial moves that would move the system out of a given basin is computational expensive (as described in refs.~\cite{xu2011direct,asenjo2014numerical,martiniani2016turning,martiniani2017numerical}), because every trial move requires an energy minimization of the $N$-body system. 
However, our approach is cheaper thaN that of refs.~\cite{xu2011direct,asenjo2014numerical,martiniani2016turning,martiniani2017numerical}, as we need not  compute the basin volume itself: we only need to perform a thermodynamic integration to compute the change of the basin free energy with temperature.

\textcolor{black}{An important feature of our approach is that all basins that we sample at high temperatures are populated with the correct Boltzmann weight at low temperatures. 
Hence, for this set of states, the concept of an effective temperature (associated with the inadequate equilibration  between basins)~\cite{cugliandolo1997energy,sciortino2005potential} does not apply.
Computational problems arise solely because some of the basins that are important at low temperatures, are hardly if ever sampled at higher temperatures.
This problem is most serious for glasses at very low temperatures, and we should expect our approach to fail in that regime. 
The degree to which  this limitation is serious can be made explicit by considering our expression for the low-temperature free energy (Eq.~\ref{eq:FL-FH}).
This equation indicates that the estimate of $F_L$ should not depend on $T_H$, and indeed this conclusion would be correct for an infinitely long simulation.
However, in any finite simulation, the sampling of low-energy ISs is likely to be inadequate. 
Hence, the dependence of our estimate of $F_L$ on $T_H$ gives an indication of the seriousness of the sampling problem. 
Such a test is shown in Fig.~\ref{KAfree} where we show the dependence of the computed equilibrium free energy of a glass for four different choice of $T_H$. 
The second point to note is that, in the absence of inadequate sampling of low-energy basins,  the equilibrium free energy of a glass cannot be protocol dependent, provided the quenching protocol cannot get stuck in local minima. 
That is why we used the simplest protocol: energy minimization. 
Of course, the free energies of the non-equilibrium glasses that have been studied in the literature, depend strongly on the protocol by which the glass has been prepared~\cite{biroli2001metastable,heuer2008exploring}.
Our approach cannot be applied to protocols based on a finite quench speed. However, in principle, the stochastic-weights method of ref.~\cite{frenkel2017monte} could be used to extend our method to compute the free energy of glasses prepared with a finite quench speed.
}

\section{Simulation Details}
To validate the method described above, we first tested it on a well-studied glassy system that can be obtained by slow (almost reversible) cooling~\cite{westergren2007silico}, namely the  Kob-Andersen (KA) binary Lennard-Jones model glass former~\cite{kob1995testing,sastry2001relationship,sengupta2011dependence}. 

We simulated $N=256$ bi-disperse spheres, 80-20 (A-B) mixture, interacting  via 
$V(r) = 4\epsilon_{\alpha\beta} \left[ \left(\frac{\sigma_{\alpha\beta}}{r}\right)^{12} - \left(\frac{\sigma_{\alpha\beta}}{r}\right)^6 \right] + 4\epsilon_{\alpha\beta} \left[c_0 + c_2 \left(\frac{r}{\sigma_{\alpha\beta}}\right)^2\right]$, for $r_{\alpha\beta} < r_c$, and zero otherwise.  
 Where $\sigma_{AA} = 1.0$, $\sigma_{AB} = 0.8$,$\sigma_{BB} = 0.88$,$r_c = 2.5*\sigma_{\alpha\beta}$, $\epsilon_{AA} = 1.0$, $\epsilon_{AB} = 1.5$, $\epsilon_{BB} = 0.5$ 
 and $r$ is the distance between the two pairs within in the cutoff distance \cite{sengupta2011dependence}. $c_0 = 0.01626656, c_2 = -0.001949974$ are correction terms to make the potential and force to go continuously to zero at cutoff. 
 In what follows, all thermodynamic quantities are expressed in reduced units: 
 $\sigma_{AA}$ is our unit of length, the unit of energy is $\epsilon_{AA}$,  $m_{A}=m_{B}=1$ is defined as the unit of mass, and the reduced temperature $T$  is expressed in units $\frac{\epsilon_{AA}}{k_B}$, where $k_B$  Boltzmann's constant. 
 Similarly,  $\beta \equiv \frac{1}{T}$. 
 
We performed NVT Monte Carlo (MC) simulations to obtain well-equilibrated  configurations at different temperatures where the system is not yet structurally arrested ($T=1-0.5$, density $\rho=N/V=1.2$). 
Below $T=0.9$,  the dynamics of the system becomes increasingly slow~\cite{sengupta2011dependence}, but equilibration of the (supercooled) liquid is still possible. 
To obtain glassy configurations \cite{ADSparmar}, we perform instantaneous  quenches from equilibrated liquid configurations at different temperatures ($T_{\text{H}}=0.5,0.6,0.7,1.0$) using the conjugate gradient minimization method \cite{press2007numerical}. 
We generate more than $500$ IS for the initial temperatures $T_{\text{H}}=0.5,0.6,0.7$ and $10^3$ configurations for quenches starting at $T_{\text{H}}=1.0$. 

Below, we report the excess free energy of the system (the ideal-gas part can be computed analytically).
For the sake of comparison, we also performed NVT MC simulations at different cooling rates.
Starting with the equilibrium liquid configurations at $T=1.0$, we perform a step-wise cooling of the system to a final temperature of $0.1$ in steps of $\Delta T = 0.1$ of duration $\Delta t$. $\Delta t$ is the amount of time the system spends at a given temperature.  In what follows, we define the cooling rate $C_{\text{r}} = \Delta T/\Delta t$.  For instance, $C_{\text{r}} = 10^{-5}$ means we perform $\Delta t = 10^4$ MC steps at a given temperature.
Each MC step involves $N$ trial displacement moves. To obtain good statistics, we performed at least $50$ independent simulation runs for two cooling rates and $N=1000$ particles.

% What we do 
\section{Basin volume method to compute free energy}
Every point (except for a set of  measure zero)  in a potential energy landscape ends up, after minimization, in one of the inherent structures. 
For this reason, it is possible to partition the configuration space of the system into basins of attraction of the different IS. 

Starting with an equilibrium liquid or supercooled liquid, we performed potential energy minimization using the CG method to obtain find the inherent structure corresponding to the initial configuration.

Strictly speaking the CG method need not yield compact basins, but in practice we observed   no noticeable effect on the calculations~\cite{asenjo2013visualizing}.
% Visualizing Basins of Attraction for Different Minimization Algorithms, D. Asenjo , J. D. Stevenson , D. J. Wales  and D. Frenkel, J. Phys. Chem. B, 117 , 12717Ð12723(2013)
In our calculation of the free energy of the quenched structures, we make use of Eq.~\ref{eq:QL_from_QH2}, which can be written as
\begin{equation}\label{eq:totf}
F(T_L)=\frac{T_L}{T_H} F(T_H)-k_B T_L \ln \left\langle e^{-[\beta_Lf^{(i)}(T_L) -\beta_Hf^{(i)}(T_H)]}\right\rangle_{MC}
\end{equation}
The difference
\begin{equation}
\beta_Lf^{(i)}(T_L) -\beta_Hf^{(i)}(T_H)\equiv \Delta \left(\frac{f^{(i)}}{T}\right)
\end{equation}
 can be computed by normal thermodynamic integration
\begin{equation}\label{eq:fbasin}
 \beta_Lf^{(i)}(T_L) = \beta_Hf^{(i)}(T_H) + \int_{\beta_H}^{\beta_L}d\beta\; \langle E^{(i)}(\beta)\rangle
\end{equation}
As there are no barriers inside a basin, this sampling should be free of hysteresis. 
In practice, we start with an IS (say, basin $i$) obtained by quenching a random equilibrium configuration at temperature $T_H$. 
We then equilibrate the system constrained to be in basin $i$, and compute its average energy. 
This step is repeated for a number of intermediate temperatures, up to $T_H$.

The steps of the algorithm are as follows:
\begin{enumerate}
\item Starting with an IS of $i-$th basin as an initial configuration, we perform MC simulations in NVT ensemble.
\item Select a particle at random and compute its energy $e_{\text{old}}$.
\item Perform a trial move by giving random displacement to the particle and compute its new energy $e_{\text{new}}$.
\item Compute the Boltzmann factor $P = \exp(-\beta (e_{\text{new}}-e_{\text{old}}))$.
\item Generate a random number $\text{rand}$. If ($\text{rand}<P$), then
\begin{enumerate}
\item If the new configuration belong to the same basin $i$, then accept the trial move.
\item Else reject the trial move and go to step $2$. 
\end{enumerate}
\item Else reject the trial move and go to step $2$.
\end{enumerate}

\subsection{Speeding up basin tests}
Sampling the properties of basins is time consuming because testing whether the system is still in the original basin~\cite{wal921}  requires us to perform a complete energy minimization for every energetically allowed MC trial move.

To reduce the computational effort required for  energy minimization, we use a structural measure to identify the basin. 
This test is based on the assumption  that a configuration belongs to the same basin of the reference IS if minimization results in a structure for which all particles have the same set of neighbors as in the IS. 
As we shall see below, this assumption is justified for the systems that we study.

To identify the nearest neighbors of a particle, we use a parameter and scale-free solid-angle based nearest-neighbor (SANN) algorithm\cite{van2012parameter}. 
Depending on the local environment of a particle, the SANN algorithm determines the set of nearest neighbours of the particle using a purely geometrical construction, similar to the Voronoi construction but much cheaper computationally~\cite{van2012parameter}. 
We stop the minimization as soon as every particle, in the configuration at the current CG step, has the same SANN neighbours as the reference IS configuration. 

As stated above, it is not obvious that configurations that have identical neighbor lists must belong to the same basin.
Therefore, we carried out extensive tests to validate this hypothesis. 
We found that the SANN identifies states that belong to the same basin as the IS for \textcolor{black}{more than $500$ IS for $T_H = 0.5,0.6,0.7$ and $10^3$ IS for $T_H=1.0$} that we tested. 
More details on the SANN-based method are given in Appendix A.

We also verified that the results of the basin sampling were independent of the minimization protocol (in a statistical sense). 
Among steepest descent, CG and FIRE methods, CG took the least time to reach the local energy minimum and hence we used the CG method to perform the potential energy minimizations.

As is clear from Eq.~\ref{eq:QL_from_QH} we need to perform a sampling over a number of basins to obtain  reliable estimates of the low-temperature  free energy. 
As we discuss in the next section, we find that a few hundred basins are usually sufficient to yield a good free-energy estimate for the KA model at moderately low temperatures.

\section{Results: KA model glass former}

To obtain glassy inherent structures,  we performed quenches of the configurations of equilibrated (supercooled) liquids at different temperatures ($T_{\text{H}}$). 
\textcolor{black}{As is well known (see e.g. ref.~\cite{saika2004distributions}), the distribution of the energies of inherent structures that are generated during a quench, depends on the initial temperature $T_{H}$ (Appendix B).
The average energy of the quenched structures is not the same as the Boltzmann-weighted average energy of these structures.\\ 
\noindent With our approach we can correctly account for the fact that the distribution over basins at low temperatures, is not the same as the ``quenched'' distribution (see Fig. 3b).
 We note that our equilibrium estimate of the average PE of an equilibrium glass generated by quenching for $T_{\text{H}} =0.5$ is significantly lower than that of  glasses generated by the slowest cooling rate that we attempted ($C_{\text{r}} = 10^{-7}$) because
\begin{equation}
\left\langle U\right\rangle_{T_L}
=\frac{\sum_{i\in {\rm sampled \; set}}U^i(T_L)\left( {q_B^i(T_L)\over q_B^i(T_H)}\right)}
{\sum_{i\in {\rm sampled \; set}} \left( {q_B^i(T_L)\over q_B^i(T_H)}\right)}
\end{equation}
The factor 
$ ( {q_B^i(T_L)/ q_B^i(T_H)})$ biases the average towards lower energies. }
\begin{figure}[h!]
%\centering
\includegraphics[scale=0.35]{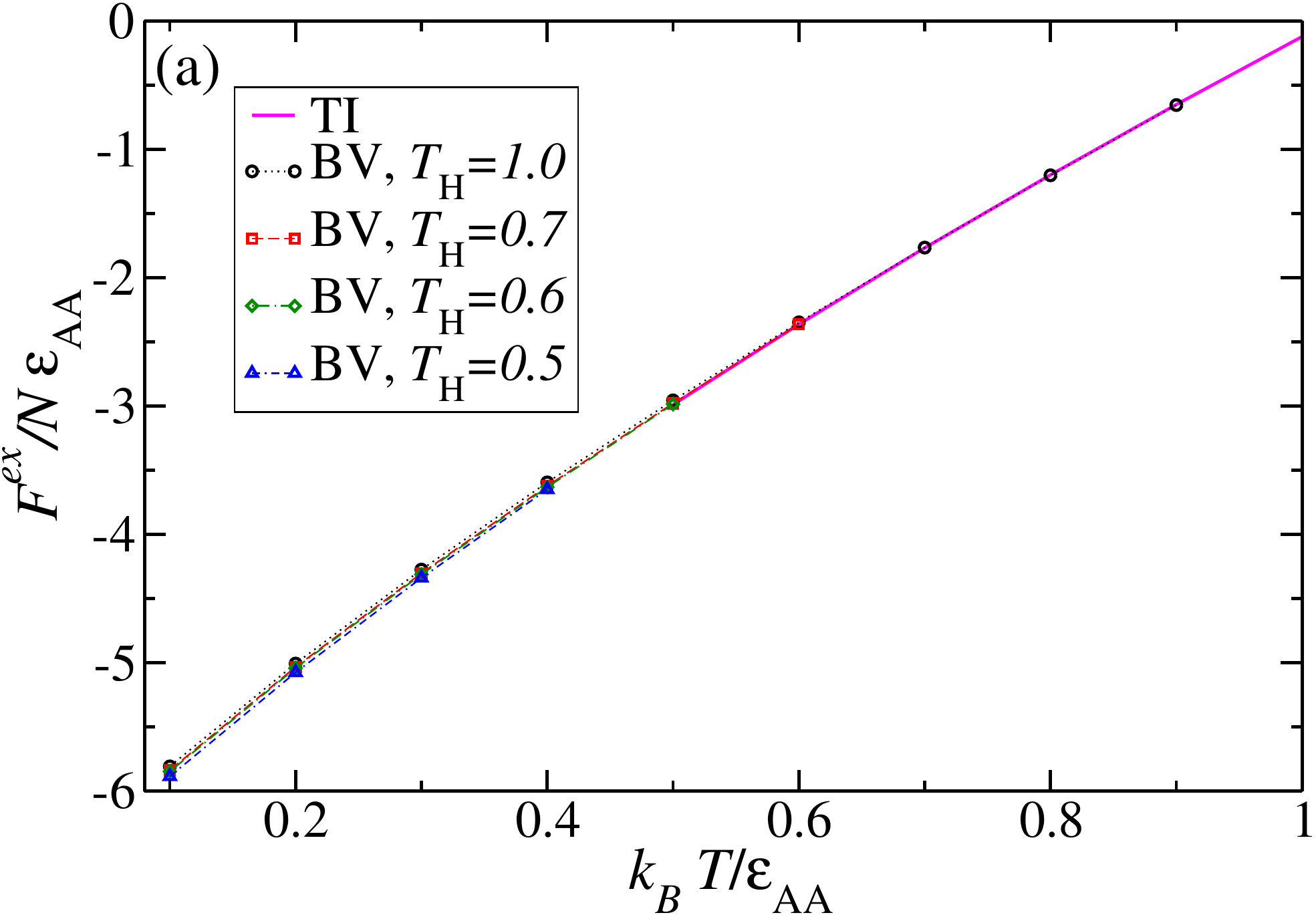}
\includegraphics[scale=0.35]{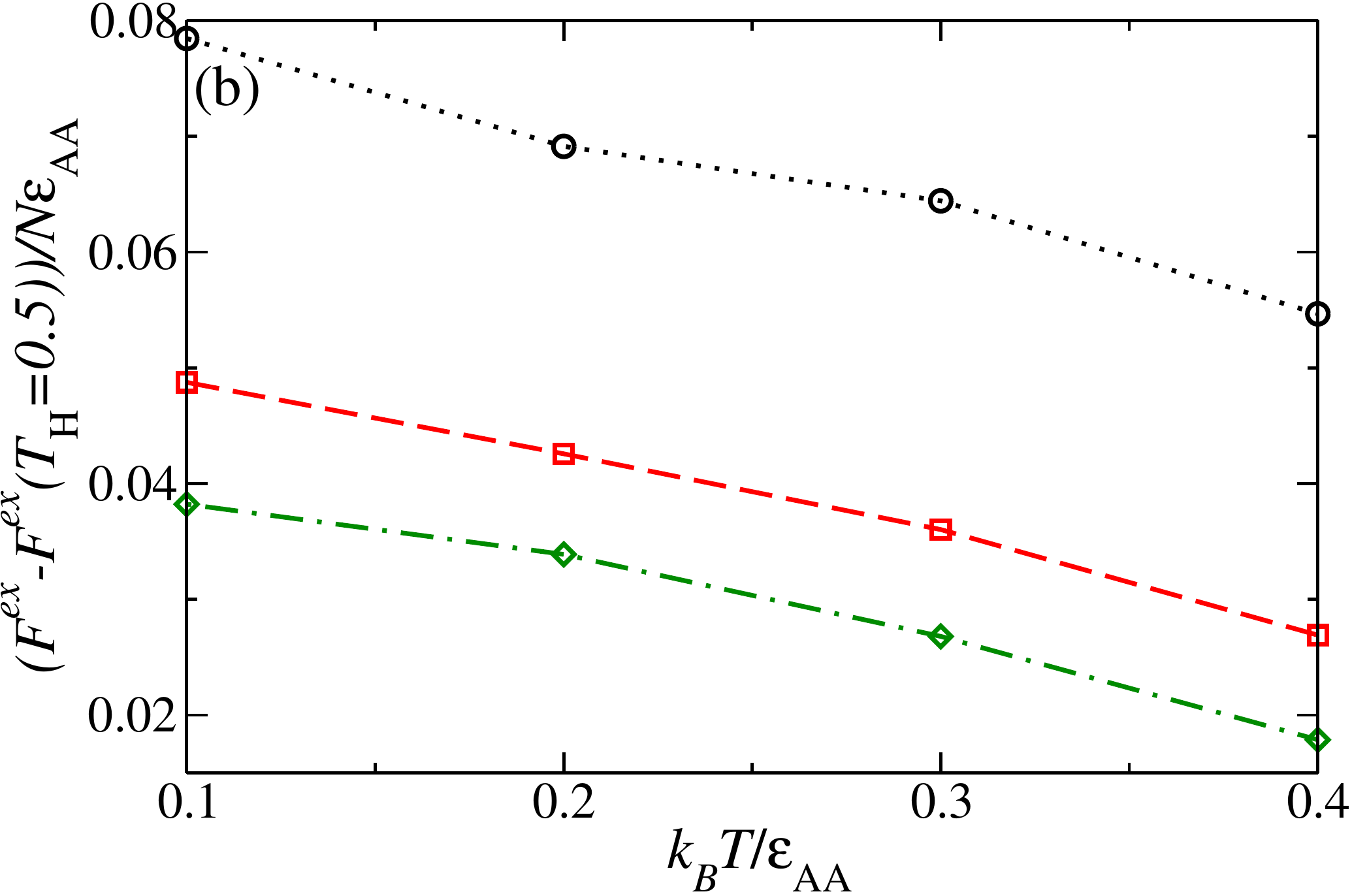}
\caption{\label{KAfree}{\bf (a)} Excess free energy per particle shown as a function of temperature, shown for different methods. 
As expected, the glasses quenched from an initial temperature $T_{\text{H}}=0.5$ are more stable than those quenched from higher $T_{\text{H}}$. 
To show the difference clearly, we subtract the excess free energy for glasses quenched from $T_{\text{H}} = 0.5$ from the free energy data with higher pre-quench temperatures $T_{\text{H}}$ in {\bf (b)}.}
\end{figure}
We use Eq. \ref{eq:fbasin} to compute the difference $\Delta \left(\frac{f^{(i)}}{T}\right)$ for each basin.  
Then, using  Eq. \ref{eq:totf}, and by averaging over different basins, we estimate the equilibrium free energy of the low-temperature glass. 

If all relevant inherent structures would have been adequately sampled at the initial temperature of the quench,  our estimate of the equilibrium free energy would not depend on $T_H$.
In Fig. \ref{KAfree}(a)(b), shows that our estimate of the free energy of the glass at lower temperatures shows a residual dependence on $T_H$. 
Although the effect is not large in the temperature range studied, it is a tell-tale sign that our high-temperature sampling misses some of the low energy inherent structures. 
We stress that this observation is not new~\cite{sastry1998signatures,saika2004distributions}.

 In Fig. \ref{KAfree}(b), we magnified the difference between different glasses by subtracting the free energy of glass with $T_{\text{H}}=0.5$ and show the free energies of different glasses. \textcolor{black}{We find that the error in the free energy estimates due to insufficient sampling of the inherent structures at high-T is non-negligible, but sufficiently small that it would have little effect on thermodynamic properties, such as solubility}. 
 In Appendix C, we show that the harmonic approximation seriously underestimates the basin free-energy difference between a low-temperature state and the same basin at
the initial temperature $T_{\text H}=1$.

\section{Conclusions}
In this paper, we have presented a novel algorithm  to estimate  the  equilibrium free energy of glasses, by performing thermodynamic integration on the basins of attraction of inherent structures that are sampled at temperatures where the fluid can equilibrate.

To speed up the sampling in the different basins, we use a structural criterion that helps us identify whether a structure belongs to the same basin as the reference IS. 
This criterion is based on the solid-angle based nearest-neighbor (SANN) algorithm of ref.~\cite{van2012parameter}. 
We have tested this method and found that it identifies the correct basin with high reliability, thereby saving  computational time. 

We have tested our method by computing free energy for the well-studied KA model glass former. 
We also study the dependence of the free energy of glasses on the preparation protocol,  specifically on the temperature of initial liquid configurations used for quenching.  \textcolor{black}{We find that, as long as the temperature of the glass is not very low, the effect on the free energy estimate due to inadequate sampling of low-energy inherent structures is small.
Our free energy estimates of the KA glass are lower than those obtained by very slow cooling.}
\textcolor{black}{Our approach could be applied to estimate the free energy of ultra-stable glasses: such information should be useful  to estimate the solubility of ultra-stable glasses.}
\\
%%%%%%%
\begin{acknowledgments}
We gratefully acknowledge the funding by the International Young Scientist Fellowship of Institute of Physics (IoP), Chinese Academy of Sciences under grant no. 2018008. We gratefully acknowledge IoP and the University of Cambridge for computational resources and support. HAV acknowledge very useful discussions with Srikanth Sastry, Jure Dobnikar, and ADS Parmar.
\end{acknowledgments}

\appendix
\section{Solid-angle based nearest-neighbor (SANN) criterion} 
\begin{figure}[h!]
%\centering
%\hspace*{-1cm}
\includegraphics[scale=0.35]{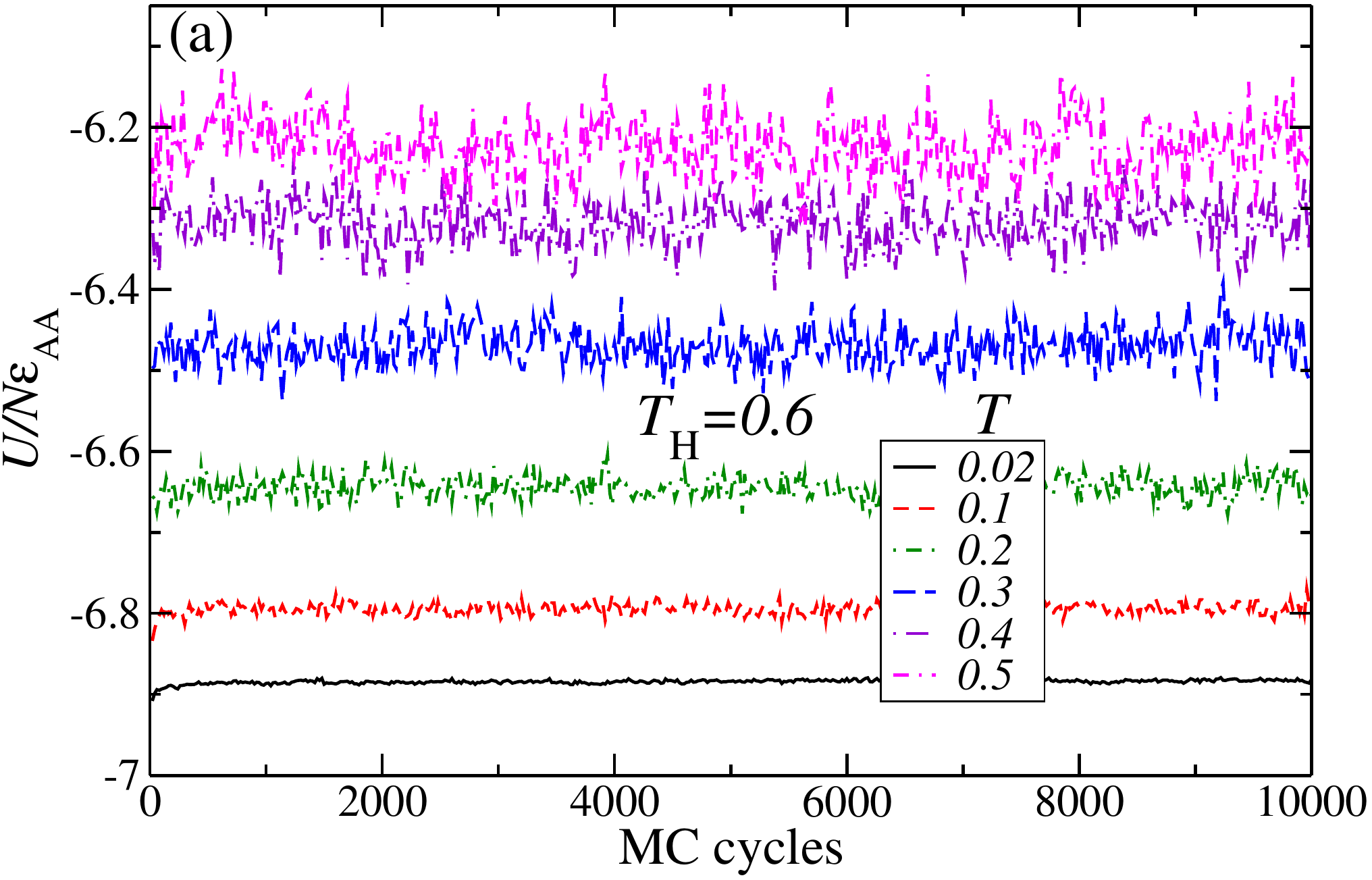}
\includegraphics[scale=0.35]{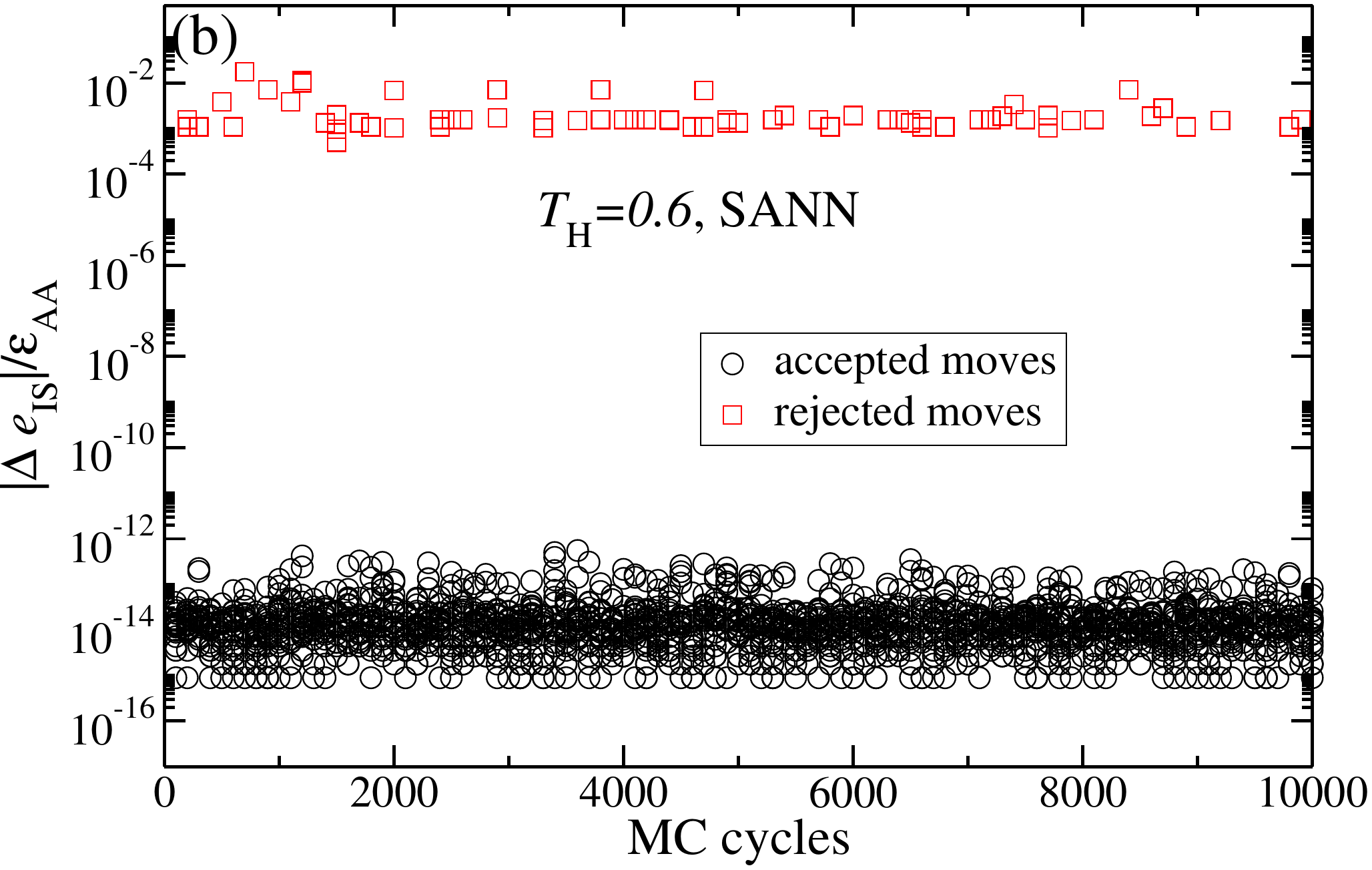}
\includegraphics[scale=0.35]{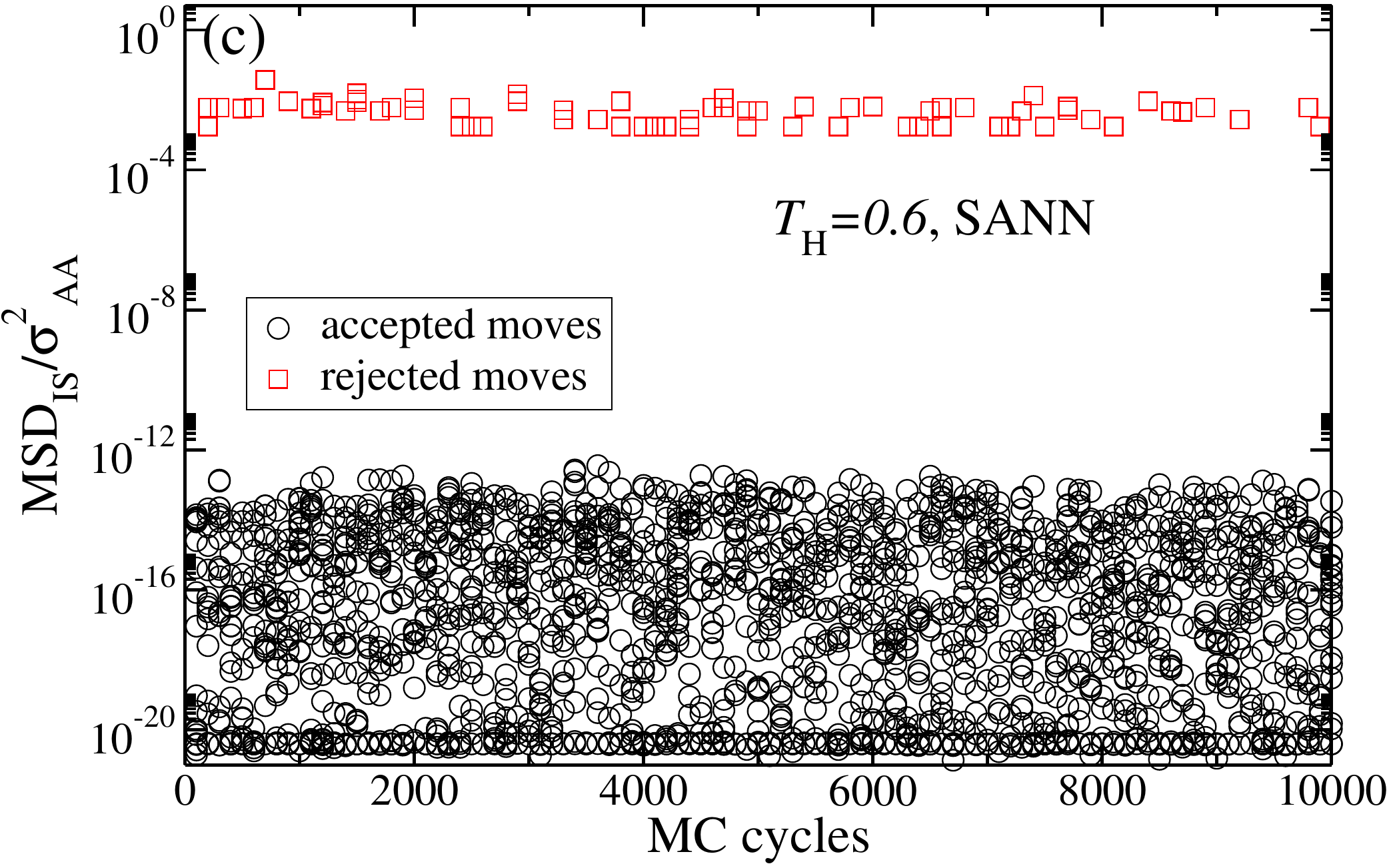}
\caption{\label{sann}{\bf (a)} The potential energy of configurations as a function of MC steps, shown for different $T$ in the BV for an IS at $T_{\text{H}}=0.6$. Observe that the system reaches equilibrium within a few hundred MC steps. {\bf (b)} The absolute value of $\Delta e_{\text{IS}}$ as a function of MC steps, for the data in {\bf(a)}. $\Delta e_{\text{IS}} = e_{\text{IS}}^{o} - e_{\text{IS}}^{\text{BV}}$, $e_{\text{IS}}^{o}$ is the per-particle inherent structure energy of the reference IS and $e_{\text{IS}}^{\text{BV}}$ are the inherent structure energies of the accepted or rejected moves in the BV method. {\bf (c)} Mean squared displacement ($\text{MSD}_{\text{IS}}$) of the reference IS and the inherent structures of the accepted or rejected moves in the BV method. Observe that the SANN criterion accurately identifies configurations that belong to the same basin ({\it i.e.,} $\Delta e_{\text{IS}}$ and $\text{MSD}_{\text{IS}}$ with values less than or equal $10^{-14}$) and those that do not.}
\end{figure}

We employ a simple structural criterion to reduce the number of CG iterations required to identify if a configuration sampled in the BV method belongs to the basin of reference IS. First, we compute the SANN list of every particle \cite{van2012parameter} in the reference IS and then during every minimization call, we decide to stop the minimization as soon as the configuration has the same SANN list as the reference IS. On average, we gain a speedup of a factor of more than $2$. In Fig. \ref{sann}, we show that the SANN method accurately identifies the basin. In Fig. \ref{sann}(a), we show the potential energy as a function of MC steps for different temperatures, shown for a single IS at $T_{\text{H}}=0.6$. We observe that the system reaches equilibrium in a few hundred MC steps. For any configuration sampled by the BV method for the reference IS at $T_{\text{H}}=0.6$, that belong to the reference basin, both the difference in the inherent structure energies and distance in the PEL, measured in terms of mean squared displacement ($\text{MSD}_{\text{IS}}$), is in the order of $10^{-14}$, see Fig. \ref{sann}(b)(c). All the rejected moves or configurations have higher energy differences and $\text{MSD}_{\text{IS}}$. We observe that the SANN criterion separates accepted moves and rejected moves, see Fig. \ref{sann}(b)(c).     
For the rejected moves, we stop the minimization when the absolute value of the energy difference between CG configurations, separated by $100$ CG iterations, reach a tolerance value of $10^{-8}$.

\section{\textcolor{black}{Distribution of inherent structure energies and the potential energy}}
\begin{figure}[h!]
%\centering
\includegraphics[scale=0.35]{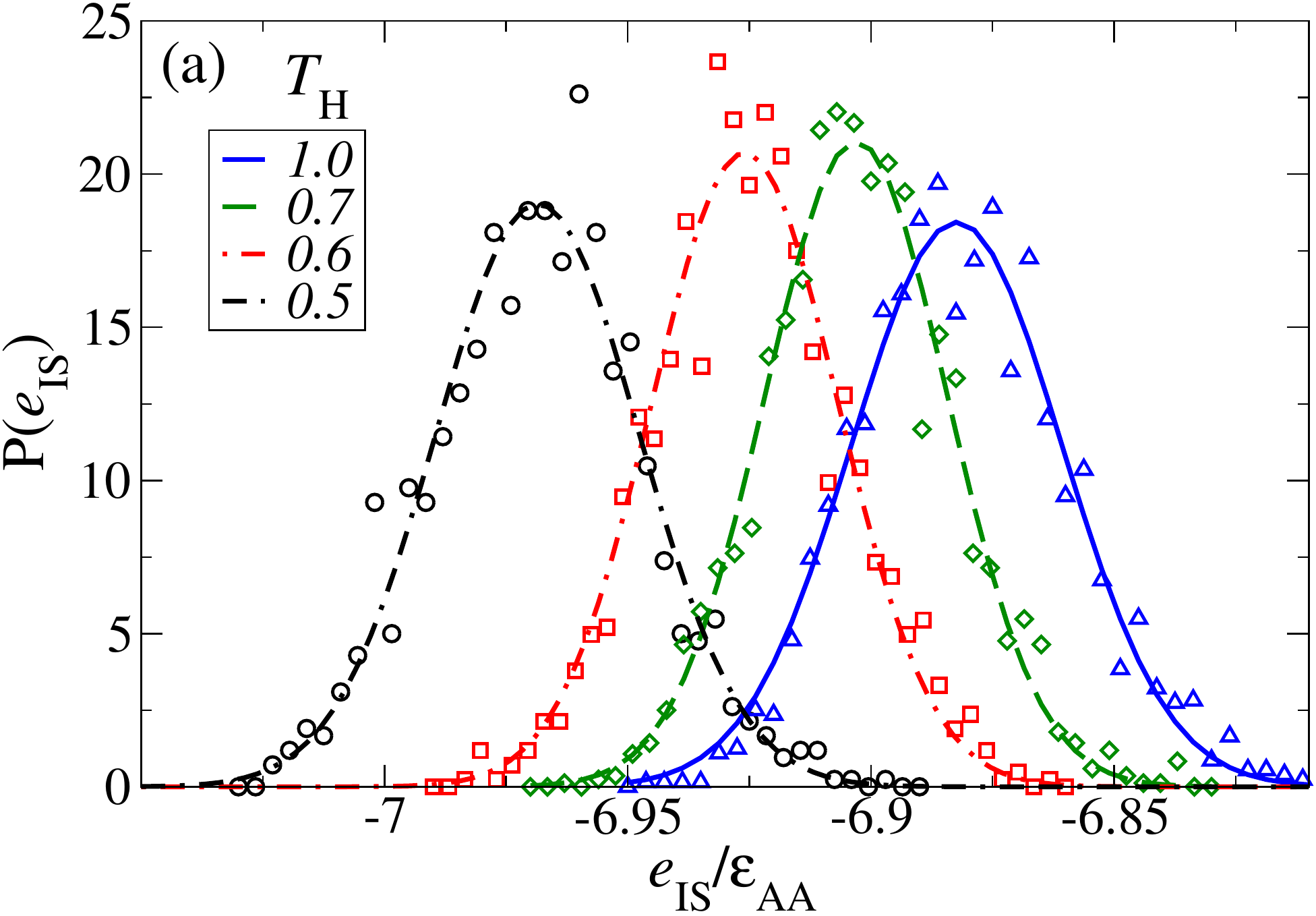}
%\hspace{1cm}
\includegraphics[scale=0.35]{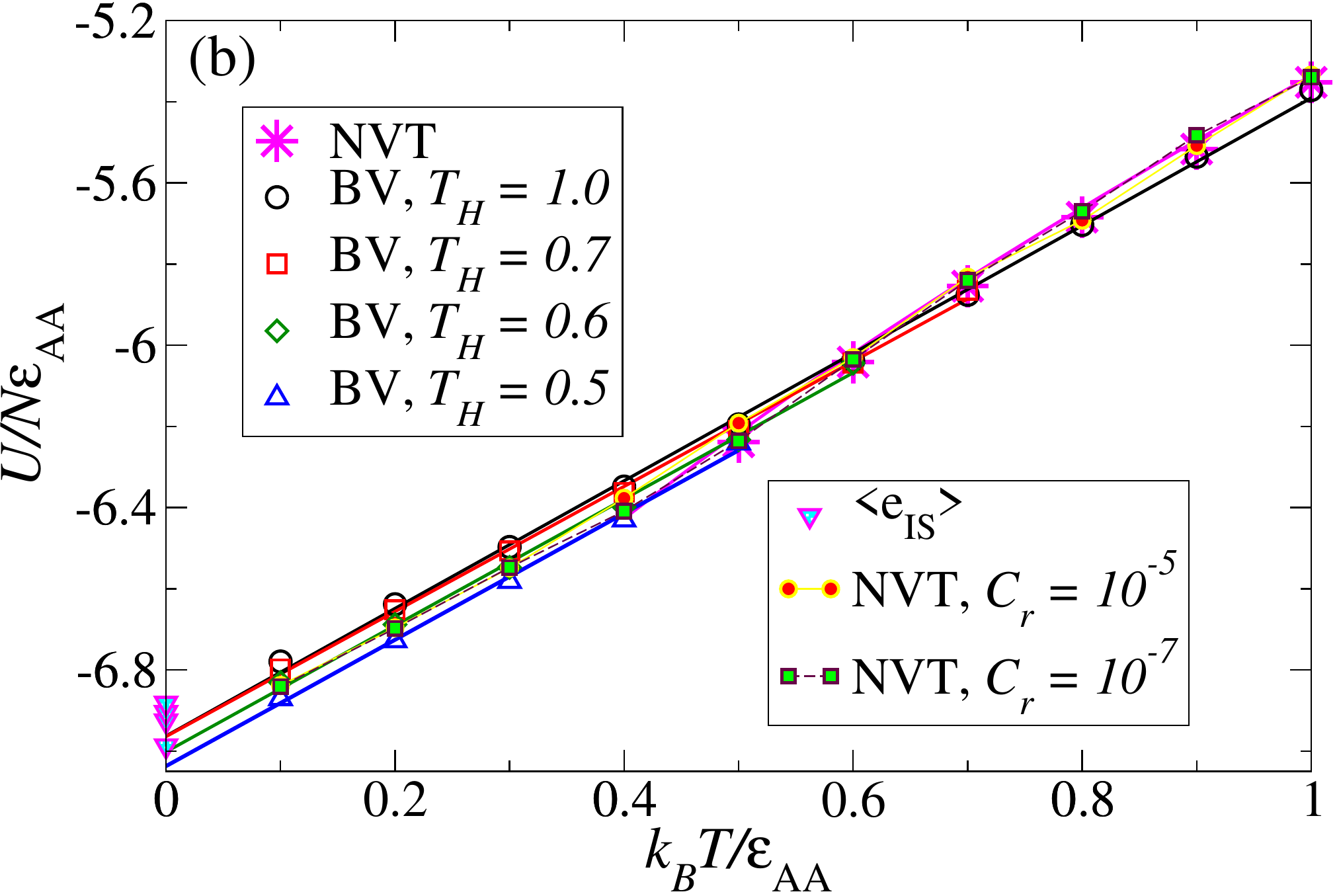}
\caption{\textcolor{black}{\label{KApe}{\bf (a)} Distributions of inherent structure energies ($e_{\text{IS}}$) that are obtained by quenching equilibrium liquid at different temperatures ($T_{\text{H}}$). 
We observe that the distributions of  $e_{\text{IS}}$ are well described by Gaussians (drawn curves) \cite{saika2004distributions}. 
The mean value of the distribution shifts to lower values with decrease in $T_{\text{H}}$.
{\bf (b)} Potential energy per particle as a function of temperature, shown for different methods. 
Note the protocol dependence of the low $T$ glasses.}}
\end{figure}

\textcolor{black}{We perform the energy minimization of equilibrated supercooled liquid configurations at different temperatures ($T_{\text{H}}$) to obtain glassy inherent structures. 
In Fig. \ref{KApe}(a), we show the distribution of inherent structure energies per particle ($e_{\text{IS}}$), obtained by quenching liquids at different $T_{\text{H}}$. \textcolor{black}{We stress that  the Gaussian form of the distributions and their dependence on $T_{H}$ is well known and has been reported in the literature \cite{saika2004distributions}}.  
The potential energy of the system at temperature $T_L$ is given by
\begin{eqnarray}\label{avpebasin}
\langle U \rangle_{T_L} &=& - \left( \frac{\partial \ln Q(T_L)}{\partial \beta_L} \right) \\
	& =& -\frac{\sum_{i \in \text{sampled set}} \left( \frac{\partial \ln q_B^i(T_L)}{\partial \beta_L} \right) \left( {q_B^i(T_L)\over q_B^i(T_H)}\right)}{\sum_{i \in \text{sampled set}} \left( {q_B^i(T_L)\over q_B^i(T_H)}\right)} \\
    &=& -\frac{\sum_{i \in \text{sampled set}} U^i(T_L) \left( {q_B^i(T_L)\over q_B^i(T_H)}\right)}{\sum_{i \in \text{sampled set}} \left( {q_B^i(T_L)\over q_B^i(T_H)}\right)}\label{avpebasin3}
\end{eqnarray}
The factor $\left( {q_B^i(T_L)\over q_B^i(T_H)}\right)$ biases the average towards the low energy states. Using the BV method, we show as a function of temperature the average potential energy (PE), computed using the above Eq. \ref{avpebasin3}, in Fig. \ref{KApe}(b). 
We compare the data from the BV method to the PE obtained for different cooling rates and NVT simulations. 
As expected, we observe that glasses obtained by quenching from $T_{\text{H}}=0.5$ have a lower energy than the glasses generated by quenching from $T_{\text{H}} = 1.0$. Observe that the PE of glasses obtained by quenching from $T_{H} =0.5$ is lower than the glasses obtained by using the lowest cooling rate ($C_{\text{r}} = 10^{-7}$).}
\section{Comparison of the basin free energy with harmonic approximation}
\begin{figure}[h!]
\centering
%\hspace*{-1cm}
\includegraphics[scale=0.35]{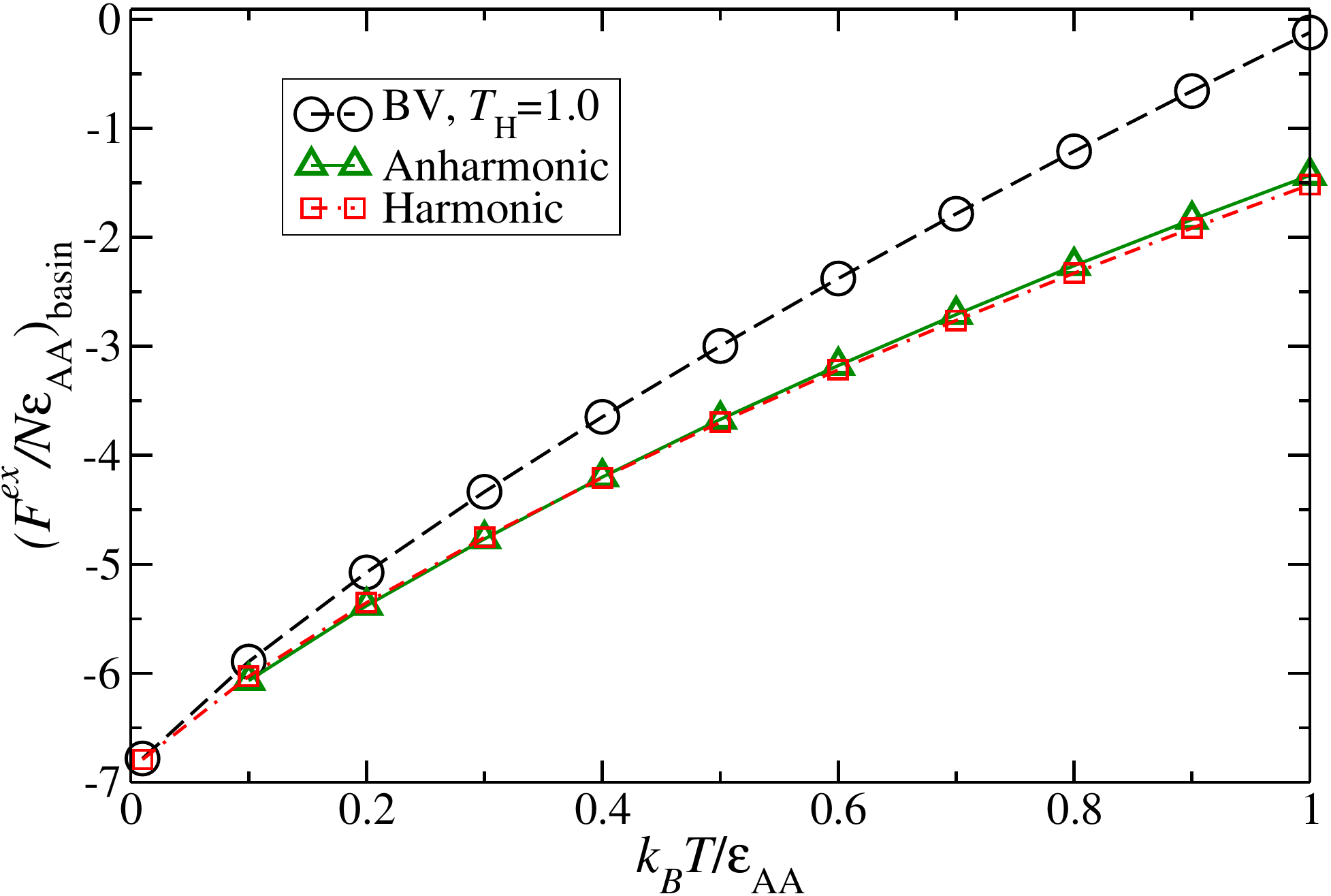}\\
%\DFcomment{I don't understand what this means. The free energy of a single basin does not depend on $T_H$}\\
\caption{\label{fbas} Comparison between the basin free energy, for a single reference IS at $T_{\text{H}}=1.0$, obtained using the BV method, the harmonic approximation and the harmonic approximation with the anharmonic corrections. Observe that the anharmonic effects are significant even at low temperatures.}
\end{figure}

The basin volume method computes the configurational free energy of glasses by computing the basin free energy of inherent structures. We can also compute the basin free energy assuming that the basins of PEL are harmonic, at low temperatures. For comparison, we compute the basin free energy of a single reference IS at $T_{\text{H}}=1.0$.  
At sufficiently low temperatures, we expect the basin is harmonic to a good approximation. Under this approximation, we can compute the canonical partition function or basin free energy of a single basin, which is given by 
\begin{equation*}
F_{\text{Harmonic}} = e_{\text{IS}} N + T \sum_{i=1}^{3N-3} \ln (\frac{\omega_i}{\sqrt{2 \pi T}}) - T\ln(V)
\end{equation*}     
Where $\omega_i^2$ are eigenvalues of the Hessian matrix at the minimum or IS.
In Fig.\ref{fbas}, we show the comparison of basin free energies, shown for the reference IS at $T_{\text{H}}=1.0$. Already at $T=0.2$, we observe a considerable difference between the basin free energy value obtained from the BV method and the harmonic approximation. This difference is due to the anharmonicity of basins. We also incorporate the anharmonic corrections to the basin free energy using the previously known methods \cite{la2003numerical,sciortino2005potential}. We observe that even with the anharmonic corrections there is no significant improvement in the estimate of the basin free energy. Therefore, the harmonic approximation, along with the anharmonic corrections, fails to give reliable free energy estimates of quenched amorphous solids.

\bibliography{free_energy_glass_v5.bib}

\end{document}